\documentclass[aip,apl,twocolumn,amsmath,amssymb,reprint]{revtex4}

\usepackage{graphicx}
\usepackage{dcolumn}
\usepackage{bm}

\begin{document}

\title{Electronic Structure and Quasiparticle Band Gap of Silicene Structures}

\author{Shouting Huang}
\affiliation{Department of Physics, Washington University in St.
Louis, St. Louis, MO 63130, USA}

\author{Wei Kang}
\affiliation{HEDPS, Center for Applied Physics and Technology, and
College of Engineering, Peking University, Beijing 100871, People
Republic of China}

\author{Li Yang}
\email{lyang@physics.wustl.edu}
\affiliation{Department of Physics, Washington University in St.
Louis, St. Louis, MO 63130, USA}

\date{\today}

\begin{abstract}
We report first-principles results on the electronic structure of
various silicene structures. For planar and simply buckled
silicenes, we confirm their zero-gap nature and show a significant
renormalization of their Fermi velocity by including many-electron
effects. However, the other two recently proposed silicene
structures exhibit a finite band gap, indicating that they are
gapped semiconductors instead of previously expected Dirac-fermion
semimetals. Moreover, our calculated quasiparticle gap
quantitatively explains the recent angle-resolved photoemission
spectroscopy measurements. In particular, the band gap of the
latter two structures can be tuned in a wide range by applying
strain, giving hope to bipolar-devices applications.
\end{abstract}

\maketitle

Graphene, a layer of hexagonal carbon lattices, has spurred tremendous
interest because of its unique linear energy-momentum dispersion and
exciting applications. \cite{2004Novoselov,2005Novoselov,2005Zhang,
2006Berger,2009Castro} Its huge success has also motivated significant
efforts to look for the similar honeycomb structures but made of
other group IV elements, such as silicon and germanium, which are
so called silicene and germanene,\cite{1994Takeda,2007Voon,
2009Cahangirov,2010Aufray,2010Padova,2010Lalmi,2010Houssa,
2011Liu,2012Vogt,2012Fleurence,2012Chen,2012Lin} respectively.
Particularly because of the compatibility with the matured silicon
technologies and the stronger spin-orbit coupling for potential
topological insulator candidates, \cite{2011Liu}
silicene is expected to be a promising material both
theoretically and practically.

Because of the strong trend of electrons of silicon atoms to form
the tetrahedral $sp^3$ hybridization, unlike graphene, a simply
buckled silicene structure is predicted to be more stable than the
perfectly planar one.\cite{2009Cahangirov} This buckling provides
a unique degree of freedom to silicene atomistic structures; to
date several unique structures with more complicated buckling
styles are proposed by recent experiments, in which a layer of
silicene has been successfully fabricated on metallic
substrates.\cite{2012Vogt,2012Fleurence,2012Chen,2012Lin} The
Dirac-fermion energy-momentum dispersion is observed in these
silicene structures as well.\cite{2012Vogt,2012Fleurence,
2012Chen,2012Lin,2012Feng} However, other than interpreting
atomistic structures, very limited attempts have been carried out
to discover the electronic structures of these recently proposed
silicene structures, which is crucial for guiding experimental
measurements and understanding the electric and optical properties
of these materials of ever-growing interest.

Moreover, the buckling of silicon atoms may break the symmetry of
the honeycomb lattice. According to the history of graphene, the
variation of structure and symmetry shall not only change the
Fermi velocity but also possibly generate a finite band gap,
promising a crucial advantage over graphene for broader
applications, such as bipolar devices and high-performance
field-effect transistors (FETs).

In this Letter, we present our first-principles calculations,
using both density functional theory (DFT) and many-body
perturbation theory (MBPT), to reveal the electronic structures of
several promising silicene structures. Our study shows that both
the planar and simply buckled silicene structures have a gapless
linear energy-momentum dispersion and that the self-energy
correction contributes to a significant enhancement of the Fermi
velocity ($\sim 37\%$). However, for the two recently proposed
silicene structures, a finite band gap is identified and this gap
value can even be varied in a wide range by tuning the strain. The
inclusion of many-electron effects by the GW approximation further
confirms this finite band gap, which corrects the previous key
interpretation of these silicene structures as massless
Dirac-fermion semimetal. At the same time, our GW calculated band
gap satisfactorily explains the recent angle-resolved
photoemission spectroscopy (ARPES) measurements. On the other
hand, the tunable band gap of silicene according to the applied
strain predicts promising device applications, overcoming the
known trouble of gapless graphene.

For our DFT simulations, both local density approximation (LDA)
and generalized-gradient approximation (GGA) are applied to make
sure our conclusions are not sensitive to the choice of
functionals. \cite{1964Hohenberg,1965Kohn} The Kohn-Sham equation
is solved using the plane-wave basis with a 24 $Ry$ energy cutoff.
Norm-conserving pseudopotentials \cite{1991Troullier} are applied
and the k-point sampling grid is set to be dense enough to obtain
converged DFT eigenvalues. In order to provide more
quantitative results with many-electron effects included, the
single-shot $G_0W_0$ approximation is employed to calculate QP
band gaps \cite{1986Hybertsen,2012Deslippe} with a layered Coulomb
truncation. The k-grid sampling of such GW calculations is
$64\times64\times1$ for planar and simply buckled silicene
structures while $32\times32\times1$ and $16\times16\times1$ for
those proposed by recent experiments, respectively.

\begin{figure}
\centering
\includegraphics[width=1.0\columnwidth]{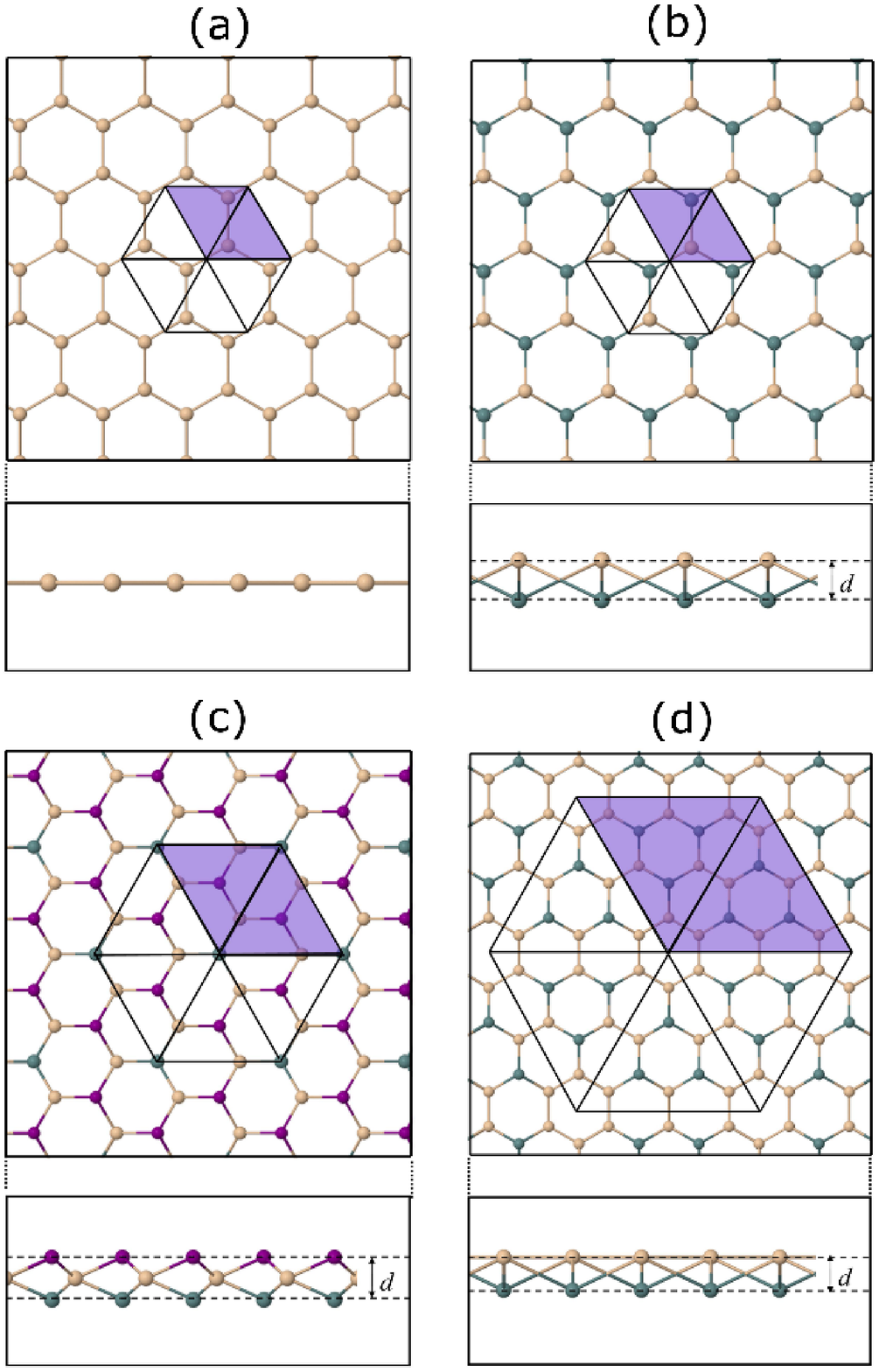}
\caption{ (Color online) The ball-stick model
of studied silicene structures. (a) the planar silicene;
(b) the simply buckled silicene; (c) buckled silicene with
a ($\sqrt3\times\sqrt3$) unit cell; (d) buckled silicene with
a ($4\times4$) unit cell. The buckling distance is labeled by $d$.
For each structure, we present both the top view and side view.
The unit cell is illustrated by the shadow region. The different
colors of silicon atoms are assigned according to their
vertical positions.
\label{fig:struc}
}
\end{figure}

Here we consider four typical silicene structures of current
fabrication interest, which are shown in Fig.~\ref{fig:struc}. The
first two are known planar and simply buckled structures while the
last two are proposed by recent experiments with ($4\times4$) and
($\sqrt3\times\sqrt3$) structures, respectively.
\cite{2012Vogt,2012Chen} The main differences between these
structures are the size of the unit cell and the way to buckle
the lattices. For example, the silicon atoms in the
($\sqrt3\times\sqrt3$) structure shown in Fig.~\ref{fig:struc} (c)
are located in three horizontal planes while those in the
($4\times4$) structure shown in Fig.~\ref{fig:struc} (d) are
located in two horizontal planes. According to the total energy of
DFT, the simply buckled structure shown in Fig.~\ref{fig:struc}
(b) is the most stable configuration with the in-plane lattice
constant of 3.81 \AA $ $ and a buckling distance $d=$0.40 \AA, which
are in agreement with previous results.
\cite{2009Cahangirov,2011Liu,2012Vogt}

It has to be pointed out that those structures shown in
Figs.~\ref{fig:struc} (c) and (d) do not exactly mimic the
experimental cases, in which the substrate may be an essential
ingredient \cite{2012Vogt,2012Chen}. This is indicated by our
DFT calculations, and other published works show that these
isolated structures are only metastable. \cite{2012Chen}
Therefore, as what have been done before,\cite{2012OHare}
our solution is to fix the buckling distance $d$ and relax
all other degrees of freedom, including force and stress.
Our following calculations will show that the essential
physical picture and main conclusions shall not be affected
by this approximation.

\begin{figure}
\centering
\includegraphics[width=1\columnwidth]{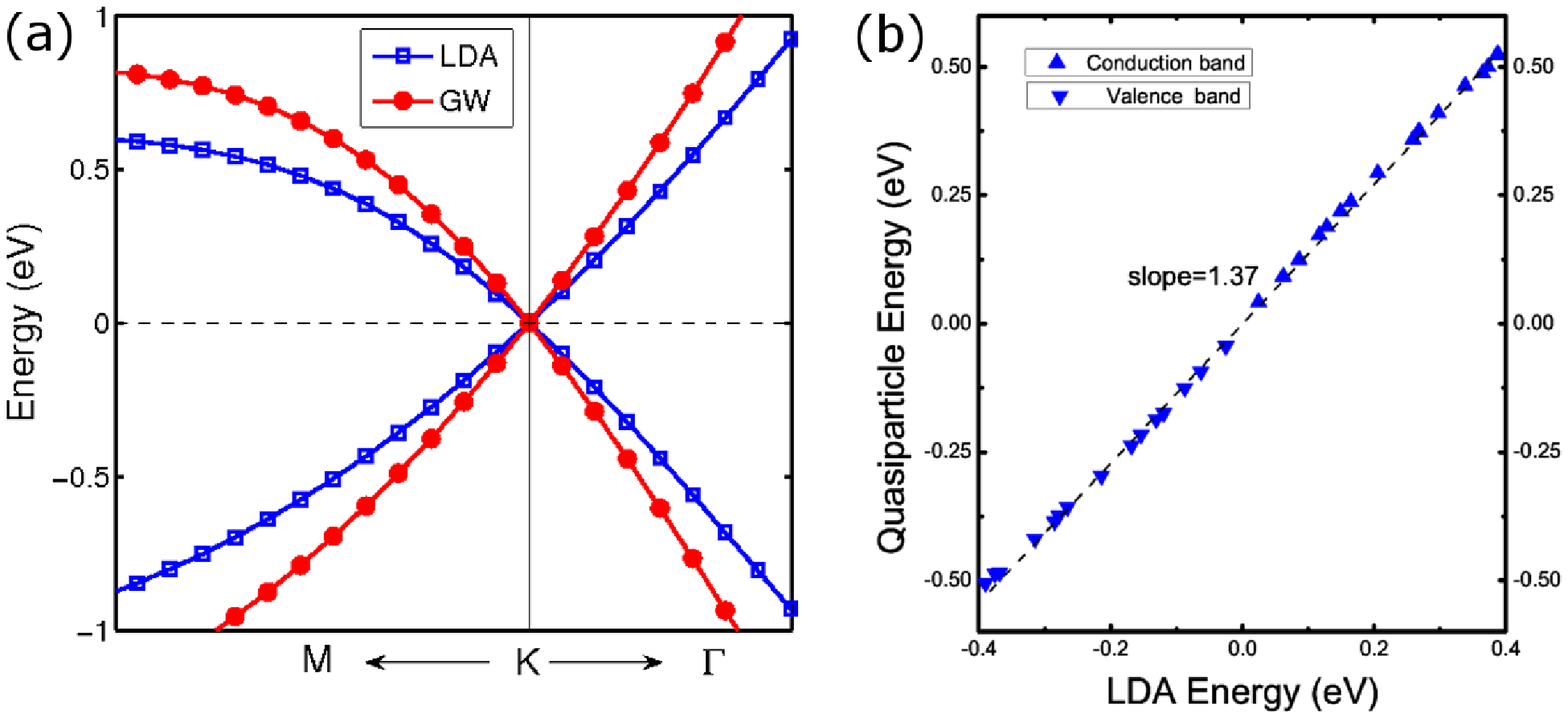}
\caption{(Color online) (a) DFT and GW calculated band structure
around the Dirac cone of simply buckled silicene. (b) The
quasiparticle energy vs. LDA band energy of simply buckled
silicene, which shows the linear correction with a slope of 1.37.}
\label{fig:band-1}
\end{figure}

For the planar and simply buckled silicene structures shown in
Figs.~\ref{fig:struc} (a) and (b), both DFT and GW calculations
confirm the zero-gap nature with a linear massless Dirac-fermion
dispersion. Since planar silicene has nearly the same
energy-momentum dispersion (with slightly different Fermi
velocities) and GW-LDA correction as the simply buckled silicene,
we only provide the results of the simply buckled case in Fig.
Fig.~\ref{fig:band-1}. Figure~\ref{fig:band-1} (a) shows the band
structures around the Dirac Cone for the simply buckled silicene,
and Fig.~\ref{fig:band-1} (b) shows its quasiparticle energy vs.
LDA band energy relation. A perfect linear relation is observed,
which means the zero-gap linear band dispersion is kept even
including many-electron effects. From these linear energy-momentum
dispersions, we get the corresponding Fermi velocity of the planar
silicene as $v_{F}=$ 5.6$\times10^{5}$ m/s and that of simple
buckled structure as $v_{F}=$ 5.4$\times10^{5}$ m/s at the DFT
level. After including the electronic self-energy corrections by
the GW approximation, these values are enhanced to be
7.7$\times10^{5}$ m/s and 7.4$\times10^{5}$ m/s, respectively.
These Fermi velocities are substantially smaller than that of
graphene ($ \sim $ 1.1$\times10^{6}$ m/s) but the renormalization
of the Fermi velocity by self-energy corrections is similar to
that of graphene, which is around a $37\%$ enhancement from their
DFT result \cite{2008Trevisanutto, 2009li} because of the
depressed screening effect.

\begin{figure}
\centering
\includegraphics[width=1\columnwidth]{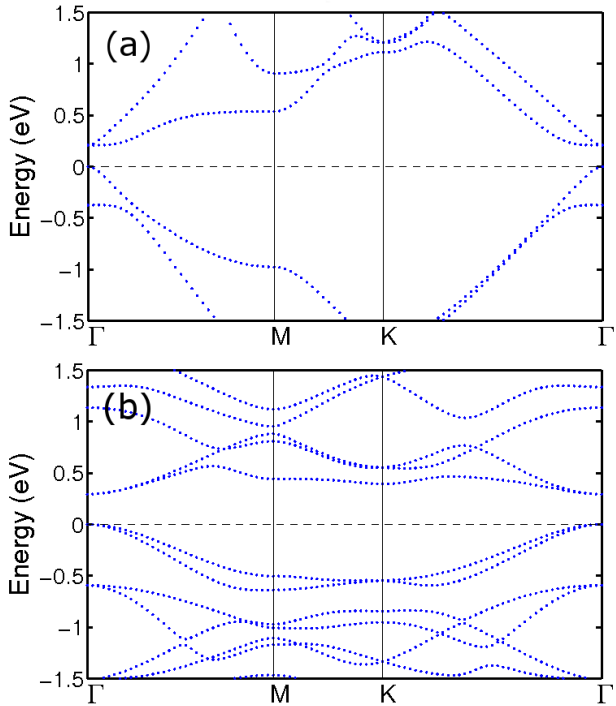}
\caption{(Color online) (a) Band structure of the
($\sqrt3\times\sqrt3$) buckled silicene structure shown
in Fig. 1 (c) with the buckling distance as $d=$ 0.71 \AA.
(b) Band structure of the ($4\times4$) buckled silicene
structure shown in Fig. 1 (d) with the buckling distance
as $d=$ 0.75 \AA.  The top of valence band is always
set to be zero.} \label{fig:band-2}
\end{figure}

Then we turn to the recently proposed structures from experiments.
In Fig.~\ref{fig:band-2} (a), the DFT-calculated band structure of
the buckled ($\sqrt3\times\sqrt3$) structure is obtained by using
the buckling distance $d=$ 0.71 \AA. which is a typical buckling
value.\cite{2009Cahangirov,2012Vogt,2012Fleurence,2012Lin}
\emph{Surprisingly, there appears to be a finite band gap ($\sim$
0.21 eV) at the $\Gamma$ point, which is qualitatively different
from previously claimed massless Dirac-fermion dispersion.}
\cite{2012Chen} Actually, we can obtain a well-defined effective
mass of the free carriers from the curvature of the band
dispersion in Fig.~\ref{fig:band-2} (a). For example, the nearly
isotropic effective mass of the light electron is 0.083 $m_0$ and
that of the hole is 0.079 $m_0$, which are typical values of
semiconductors.

\begin{figure}
\centering
\includegraphics[width=1\columnwidth]{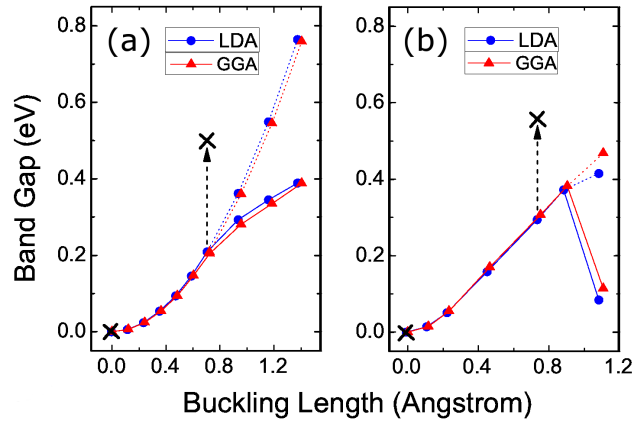}
\caption{(Color online) The band gap evolution as the variation
of buckling distance $d$ for (a) the ($\sqrt3\times\sqrt3$)
silicene structure and (b) the ($4\times4$) silicene structure.
The direct band gap values are connected by dash lines while
the indirect band gap values are connected by solid lines.
The GW results are marked by cross signs.
} \label{fig:gap}
\end{figure}

Moreover, we further tune the buckling distance $d$ while
maintaining the supercell geometry to obtain the evolution of the
band gap, which is presented in Fig.~\ref{fig:gap} (a). From
another point of view, this is equivalent to tune the strain
condition of the layer structure. For both LDA and GGA results,
the band gap of such a ($\sqrt3\times\sqrt3$) structure increases
while enlarging the buckling distance. In particular, as the
buckling distance is more than 0.8 \AA, we see a transition of the
band gap from the direct one to the indirect one, which is
reasonable because the band structure shall approach that of
tetrahedral silicon due to the stronger $sp^3$ hybridization.
Therefore, unlike the planar and simply buckled silicene, whose
band gaps are always zero under uniaxial or biaxial strain, the
band gap of this ($\sqrt3\times\sqrt3$) silicene structure can be
tuned from zero to a wide range by the mechanical strain, which
shall be of broad interests.

To further confirm this finite band-gap nature, we have performed
the GW calculation. For the purpose of justification, one GW
calculation of a typical buckling distance shall be enough;
as shown in Fig.~\ref{fig:gap} (a), the QP band gap is around
0.5 eV for the $d=$ 0.71 \AA $ $ buckled case, which is almost a
140$\%$ enhancement from the DFT result because of the substantially
depressed screening. It must be addressed again that our referenced
experiments use a metallic substrate, which may reduce the self-energy
correction due to the metallic screening and possible charge transfer.
However, all of these factors will not close the finite band gap and,
usually, the realistic QP band gap shall still be slightly larger
than the DFT result.\cite{2006neaton,2011li}

We now turn to the band structure of the other important silicene
structure ($4\times4$), as shown in Fig.~\ref{fig:band-2} (b). For
this superstructure, the buckling distance $d=$ 0.75 \AA $ $ is
chosen from the experimental study. \cite{2012Vogt}
 This time we observe a finite band gap again,
which is a typical direct-band-gap semiconductor as shown in
Fig.~\ref{fig:band-2} (b). The evolution of the band gap with
the buckling distance d is also presented in Fig.~\ref{fig:gap} (b).
Similar to the ($\sqrt3\times\sqrt3$) case, the band gap increases
as we enlarge the buckling distance. When the buckling distance
is more than $d=$ 0.8 \AA, the similar direct-to-indirect band
gap transition is observed in Fig.~\ref{fig:gap} (b). The
corresponding GW results are marked as well, predicting a
larger QP band gap than the DFT results.

Interestingly, when we consult the relevant experimental result,
a finite band gap was observed, but it had been attributed to
extrinsic factors, such as the layer-substrate interaction.
\cite{2012Vogt} However, our simulation shows that this finite
band gap may be intrinsic if the silicene sample possesses the
claimed ($4\times4$) structure. Moreover, if we compare the
experimentally observed band gap value by ARPES to our
first-principles GW result shown in Fig.~\ref{fig:gap} (b),
they are very close to each other, 0.56 eV from our calculation
while 0.6 eV from experimental work, for the ($4\times4$)
structure with a buckling distance of $d=$ 0.75 \AA. \cite{2012Vogt}

After presenting first-principles results, it is necessary to
figure out the physical reason for such a finite band gap
opening in these promising silicene structures. According
to past intensive studies on graphene, a necessary condition
for the massless Dirac cone with a zero-gap character is the
inversion symmetry of AB sublattices. However, for the
structures shown in Figs.~\ref{fig:struc} (c) and (d),
A and B sublattices are no longer equivalent to each other,
resulting in the broken inversion symmetry and, thereafter,
generating a finite band gap. A similar idea was noticed
before, \emph{e.g.}, the applied gate electric field can
break this symmetry and introduced a finite band gap in
simply buckled silicene and germanene.
\cite{2012Ni,2012OHare,2012falko}.

\begin{figure}
\centering
\includegraphics[width=1\columnwidth]{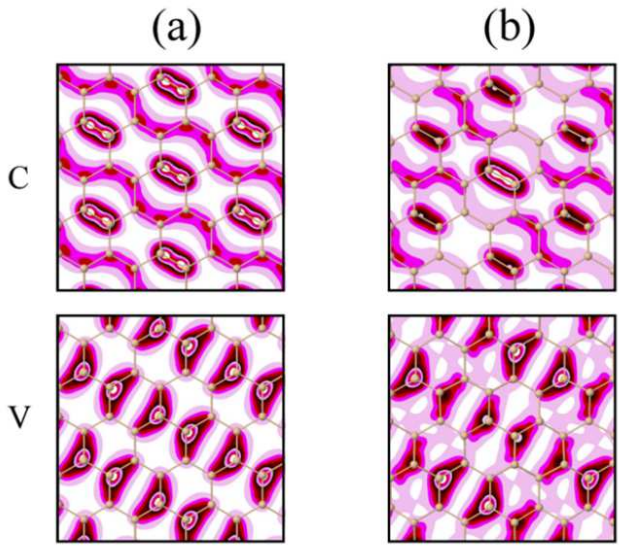}
\caption{(Color online) Top view of color contour plots
of the charge distribution of typical conduction and
valence states at energy extrema around the band gap
for (a) the planar silicene when $d=$ 0 and (b) the
($4\times4$) silicene structure when $d=$ 0.75 \AA.
} \label{fig:charge}
\end{figure}

In order to better see this broken AB sublattice symmetry,
we present the charge distributions of valence and conduction
states at energy extrema around the band gap. As shown in
Fig.~\ref{fig:charge} (a), for planar silicene, we observe
the charge distributions of both valence and conduction
states are nearly identical for A and B sublattices.
However, for the ($4\times4$) silicene structure,
we can clearly see the inequivalent charge distribution
of A and B sublattices as shown in Fig.~\ref{fig:charge} (b)
due to the broken symmetry.

All the above calculations consider only free-standing silicene
structures. However, the predicted finite band-gaps in the
structures, as shown in Figs.~\ref{fig:struc} (c) and (d), will
persist even if the substrate effect is included because the
layer-substrate interactions shall further break the symmetry and
enlarge the band gap. \cite{2007Zhou} In this sense, we believe
that the recently proposed silicene structures have qualitatively
different electronic structures and band gaps from the widely
assumed massless Dirac-fermion dispersion. The most likely reason
for this conflict between our simulation results and previous
experimental conclusions may be due to the interpretations of
atomic structures from their experimental data.

Finally we have to point out that the finite-gap nature of these
recently proposed silicene structures shall be of practical
interest. As shown in  Figs.~\ref{fig:band-2} and ~\ref{fig:gap},
the band gap and even the band topology can be tuned in a wide
range by the buckling distance $d$ and the associated strain; this
provides a precious degree of freedom to control the electronic
and optical properties of such materials, making them superior to
gapless graphene. According to previous studies,
\cite{2009Cahangirov,2010Lalmi,2012Vogt,2012Fleurence,2012Lin} the
lattice structure and buckling distance are able to be varied by
layer-substrate interactions. Therefore, if future experiments can
fabricate and confirm more similar structures, they shall be of
broad interest for devices applications with such a tunable band
gap.

In conclusion, we employ first-principles simulations to
study the electronic structure of various silicenes.
Our results show that the band structure of silicene is
sensitive to the buckled atomistic structure. Unlike the
planar and simply buckled ones, these silicene structures
proposed by recent experiments exhibit a finite band gap,
making them promising 2D semiconductors instead of zero-gap
semimetals, as were previously assumed. On the other hand,
this finite band gap can be tuned in a wide range by strains
and makes these silicene structures superior to graphene
for many important applications. Our predictions reinterpret
relevant experimental measurements and shall motivate
more reliable justifications.

We thank Ryan Soklaski for proofreading. S.H. and L.Y. acknowledge
the support by NSF Grant No. DMR-1207141 and the International
Center for Advanced Renewable Energy and Sustainability (I-CARES)
of the Washington University. The computational resources have
been provided by Lonestar of Teragrid at the Texas Advanced
Computing Center and Hoppers of the National Energy Research
Scientific Computing Center (NERSC). The ground state calculation
is performed by the Quantum Espresso\cite{2009Giannozzi}. The GW
calculation is done with the BerkeleyGW
package\cite{2012Deslippe}.


%
%

%




\begin{thebibliography}{99}

\bibitem{2004Novoselov}
K. S. Novoselov, A. K. Geim, S. V. Morozov, D. Jiang, Y. Zhang, S. V. Dubonos,
I. V. Grigorieva, and A. A. Firsov, Science \textbf{306}, 666 (2004).

\bibitem{2005Novoselov}
K. S. Novoselov, A. K. Geim, S. V. Morozov, D. Jiang, M. I. Katsnelson,
I. V. Grigorieva, S. V. Dubonos, and A. A. Firsov, Nature (London) \textbf{438}, 197 (2005).

\bibitem{2005Zhang}
Y. Zhang, Y.-W. Tan, H. L. Stormer, and P. Kim, Nature (London) \textbf{438}, 201 (2005).

\bibitem{2006Berger}
C. Berger, Z. Song, X. Li, X. Wu, N. Brown, C. Naud, D. Mayou,
T. Li, J. Hass, A. N. Marchenkov, E. H. Conrad, P. N. First, and
W. A. de Heer, Science \textbf{312}, 1191 (2006).

\bibitem{2009Castro}
A. H. Castro Neto, F. Guinea, N. M. R. Peres, K. S. Novoselov, and A. K. Geim,
Rev. Mod. Phys. \textbf{81}, 109 (2009).

\bibitem{1994Takeda}
K. Takeda and K. Shiraishi, Phys. Rev. B \textbf{50}, 14916 (1994).

\bibitem{2007Voon}
G. G. Guzm\'{a}n-Verri and L. C. L. Voon, Phys. Rev. B \textbf{76}, 075131 (2007).

\bibitem{2009Cahangirov}
S. Cahangirov, M. Topsakal, E. Akt\"urk, H. $\ifmmode \mbox{\c{S}}\else \c{S}\fi{}$ahin, and S. Ciraci, Phys. Rev. Lett. \textbf{102}, 236804 (2009).

\bibitem{2010Aufray}
B. Aufray, A. Kara, S. Vizzini, H. Oughaddou, C. L\'{e}andri,
B. Ealet, and G. L. Lay, Appl. Phys. Lett. \textbf{96}, 183102 (2010).

\bibitem{2010Padova}
P. D. Padova, C. Quaresima, C. Ottaviani, P. M. Sheverdyaeva,
P. Moras, C. Carbone, D. Topwal, B. Olivieri, A. Kara,
H. Oughaddou, B. Aufray, and G. L. Lay, Appl. Phys. Lett. \textbf{96}, 261905 (2010).

\bibitem{2010Lalmi}
B. Lalmi, H. Oughaddou, H. Enriquez, A. Kara, S. Vizzini,
B. Ealet, and B. Aufray, Appl. Phys. Lett. \textbf{97}, 223109 (2010).

\bibitem{2010Houssa}
M. Houssa, G. Pourtois, V. V. Afanas'ev, and A. Stesmans, Appl. Phys. Lett. \textbf{97}, 112106 (2010).

\bibitem{2011Liu}
C.-C. Liu, W. Feng, and Y. Yao, Phys. Rev. Lett. \textbf{107}, 076802 (2011).

\bibitem{2012Vogt}
P. Vogt, P. De Padova, C. Quaresima, J. Avila, E. Frantzeskakis,
M. C. Asensio, A. Resta, B. Ealet, and G. Le Lay,
Phys. Rev. Lett. \textbf{108}, 155501 (2012).

\bibitem{2012Fleurence}
A. Fleurence, R. Friedlein, T. Ozaki, H. Kawai,
Y. Wang, and Y. Yamada-Takamura,
Phys. Rev. Lett. \textbf{108}, 245501 (2012).

\bibitem{2012Chen}
L. Chen, C.-C. Liu, B. Feng, X. He, P. Cheng, Z. Ding, S. Meng,
Y. Yao, and K. Wu, Phys. Rev. Lett. \textbf{109}, 056804 (2012).

\bibitem{2012Lin}
C.-L. Lin, R. Arafune, K. Kawahara, N. Tsukahara, E. Minamitani,
Y. Kim, N. Takagi, and M. Kawai,
Appl. Phys. Express \textbf{5}, 045802 (2012).

\bibitem{2012Feng}
B. Feng, Z. Ding, S. Meng, Y. Yao, X. He, P. Cheng, L. Chen,
and K. Wu, Nano Lett., \textbf{12}, 3507 (2012).

\bibitem{1964Hohenberg}
P. Hohenberg and W. Kohn, Phys. Rev. \textbf{136}, 864 (1964).

\bibitem{1965Kohn}
W. Kohn and L. J. Sham, Phys. Rev. \textbf{140}, 1133 (1965).

\bibitem{1991Troullier}
N. Troullier and J. L. Martins, Phys. Rev. B \textbf{43}, 1993 (1991).

\bibitem{1986Hybertsen}
M. S. Hybertsen and S. G. Louie, Phys. Rev. B \textbf{34}, 5390 (1986).

\bibitem{2012Deslippe}
J. Deslippe, G. Samsonidze, D. A. Strubbe, M. Jain, M. L. Cohen, and S. G. Louie,
Comput. Phys. Commun. \textbf{183}, 1269 (2012).

\bibitem{2008Trevisanutto}
P.E. Trevisanutto, C. Giorgetti, L. Reining, M. Ladisa, and V. Olevano, Phys. Rev. Lett. \textbf{101}, 226405 (2008).

\bibitem{2009li}
L. Yang, J. Deslippe, C.-H. Park, M. L. Cohen, and S. G. Louie,
Phys. Rev. Lett. \textbf{103}, 186102 (2009).

\bibitem{2006neaton}
J. B. Neaton, M. S. Hybertsen and S. G. Louie, Phys. Rev. Lett.
\textbf{97}, 216405 (2006).

\bibitem{2011li}
L. Yang, Nano Lett. \textbf{11}, 3844 (2011).

\bibitem{2012Ni}
Z. Ni, Q. Liu, K. Tang, J. Zheng, J. Zhou, R. Qin, Z. Gao, D. Yu,
and J. Lu,, Nano Lett. \textbf{12}, 113 (2012).

\bibitem{2012OHare}
A. O'Hare, F. V. Kusmartsev, and K. I. Kugel, Nano Lett. \textbf{12}, 1045 (2012).

\bibitem{2012falko}
N. D. Drummond, V. Z\'olyomi, and V. I. Fal'ko,
Phys. Rev. B \textbf{85}, 075423 (2012).

\bibitem{2007Zhou}
S. Y. Zhou, G.-H. Gweon, A. V. Fedorov, P. N. First, W. A.
de Heer, D.-H. Lee, F. Guinea, A. H. Castro Neto, and A. Lanzara,
Nature Mater. \textbf{6}, 770 (2007).

\bibitem{2009Giannozzi}
P. Giannozzi \textit{et al}., J.Phys.:Condens.Matter, \textbf{21}, 395502 (2009).

\end{thebibliography}


\end{document}